\documentclass[11pt]{article}
\usepackage[textwidth=15.2cm,textheight=22cm]{geometry}
\usepackage{amsmath,amssymb}
\usepackage{latexsym}
\usepackage{multicol}
\usepackage{graphicx}
\usepackage{bm}
\allowdisplaybreaks[1]

\newcommand{\del}{\partial}
\newcommand{\be}{\begin{equation}}
\newcommand{\ee}{\end{equation}}
\newcommand{\ba}{\begin{eqnarray}}
\newcommand{\ea}{\end{eqnarray}}
\newcommand{\bdm}{\begin{displaymath}}
\newcommand{\edm}{\end{displaymath}}

\newcommand\fr[1]{\frac{1}{#1}}

\newcommand{\rom}[1]{\uppercase\expandafter{\romannumeral #1\relax}}
\newcommand{\nbar}[1]{\overline{#1}}

\def\ba{\bar A}

\def\beq{\begin{equation}}
\def\eeq{\end{equation}}
\newcommand{\half}{\frac{1}{2}}
\newcommand{\nn}{\nonumber}

\newcommand{\ndt}{\noindent}

\newcommand{\D}{\mathcal{D}}
\newcommand{\bD}{\bar{\mathcal{D}}}
\def\bea{\begin{eqnarray}}
\def\eea{\end{eqnarray}}
\def\beas{\begin{eqnarray*}}
\def\eeas{\end{eqnarray*}}
\def\sla{\raise.15ex\hbox{$/$}\kern-.57em}

\def\parm{{\partial}_{-}}

\def\spa#1.#2{\left\langle#1\,#2\right\rangle}
\def\spb#1.#2{\left[#1\,#2\right]}

\begin{document}

\begin{titlepage}
\begin{flushright}    
{\small $\,$}
\end{flushright}
\vskip 1cm
\centerline{\Large{\bf{Gravitation and}}}
\vskip 0.5cm
\centerline{\Large{\bf{quadratic forms}}}
\vskip 1.5cm
\centerline{Sudarshan Ananth$^\dagger$, Lars Brink$^{\;\sharp}$$^{\;*}$, Sucheta Majumdar$^\dagger$, Mahendra Mali$^\diamond$ and Nabha Shah$^\dagger$}
\vskip 1cm
\centerline{${\,}^\dagger$\it {Indian Institute of Science Education and Research}}
\centerline{\it {Pune 411008, India}}
\vskip 1cm
\centerline{${\,}^\sharp$\it {Department of Physics}}
\centerline {\it {Chalmers University of Technology, S-41296 G\"oteborg, Sweden}}
\vskip 1cm
\centerline{${\,}^*$\it {Institute of Advanced Studies and Department of Physics \& Applied Physics}}
\centerline {\it {Nanyang Technological University, Singapore 637371, Singapore}}
\vskip 1cm
\centerline{${\,}^\diamond$\it {School of Physics, Indian Institute of Science Education and Research}}
\centerline{\it {Thiruvananthapuram, Trivandrum 695016, India}}
\vskip 1.5cm
\centerline{\bf {Abstract}}
\vskip .5cm
The light-cone Hamiltonians describing both pure ($\mathcal N=0$) Yang-Mills and $\mathcal N=4$ super Yang-Mills may be expressed as quadratic forms. Here, we show that this feature extends to theories of gravity. We demonstrate how the Hamiltonians of both pure gravity and $\mathcal N=8$ supergravity, in four dimensions, may be written as quadratic forms. We examine the effect of residual reparametrizations on the Hamiltonian and the resulting quadratic form. 
\vfill
\end{titlepage}

\section{Introduction}

\ndt In a recent paper~\cite{ABM} we showed that the light-cone gauge Hamiltonians describing pure Yang-Mills theory and $\mathcal N=4$ Yang-Mills theory could be expressed as quadratic forms. Interestingly, this quadratic form structure occurs exclusively in the $\mathcal N=0$ (non-supersymmetric) and $\mathcal N=4$ (maximally supersymmetric) cases. In this paper, we extend our analysis to theories of gravity and show that the same holds true for both pure gravity and the maximally supersymmetric $\mathcal N=8$ supergravity in four dimensions. Simple mathematical structures are often signatures of a hidden symmetry and this makes them interesting. There is growing evidence that pure gravity in $d=4$ may have hidden symmetries~\cite{MH} and that some of the surprising ultraviolet cancelations encountered in the $\mathcal N=8$ theory may originate from pure gravity itself~\cite{ZB2} - unexpected cancelations themselves being another reliable indicator of hidden symmetries. We hope that the quadratic form structures introduced in this paper are signatures of such a symmetry, with possible links to the recent work of~\cite{BMS}. 
\vskip 0.3cm
\ndt Apart from their Hamiltonians being quadratic forms, Yang-Mills theory and Gravity share other close links including the KLT relations~\cite{KLT}. These relations seem to suggest that the finiteness properties of $\mathcal N=4$ super Yang-Mills theory could possibly carry over to $\mathcal N=8$ supergravity~\cite{ZB1}. An open question in this regard is how much of the unique quantum behaviour of maximally supersymmetric theories is due entirely to supersymmetry (in the $\mathcal N=8$ theory, the exceptional symmetry plays an equally~\cite{ABMj} important role).
\vskip 0.3cm
\ndt Our focus in this paper is on the Hamiltonians describing pure gravity and $\mathcal N=8$ supergravity in light-cone gauge. In the next section, we start by reviewing the formulation of pure gravity, in $d=4$, in light-cone gauge. This leads to a closed-form expression for the Lagrangian based entirely on the physical fields in the theory. From the closed-form result, we extract the kinetic term and the higher-point interaction vertices. We write down the corresponding Hamiltonian and describe both its residual symmetry and the quadratic form structure it possesses to order $\kappa$. In section $3$, we study the Hamiltonian to order $\kappa^2$ and show that it may be expressed as a quadratic form. We prove its invariance under the residual reparametrization symmetry. Finally, we briefly review how the quadratic form also appears in $\mathcal N=8$ supergravity.

\vskip 0.3cm

\section{Pure Gravity in $d=4$}

\ndt With the metric $(-,+,+,+)$, the light-cone coordinates are
\bea
x^\pm=\fr{\sqrt 2}(x^0\pm x^3)\ ,
\eea
with the corresponding derivatives $\partial_\pm$. The transverse coordinates and derivatives are
\bea
x=\frac{1}{\sqrt 2}\,(\,{x_1}\,+\,i\,{x_2}\,)\ ;\qquad  {\bar\partial} =\frac{1}{\sqrt 2}\,(\,{\partial_1}\,-\,i\,{\partial_2}\,)\ .
\eea
\ndt On a Minkowski background, where the cosmological constant $\Lambda$ vanishes, the Einstein-Hilbert action reads 
\bea
S_{EH}=\int\,{d^4}x\;\mathcal{L}\,=\,\frac{1}{2\,\kappa^2}\,\int\,{d^4}x\;{\sqrt {-g}}\,\,{\mathcal R}\ ,
\eea
where $g=det\,(\,{g_{\mu\nu}}\,)$ is the determinant of the dynamical variable, the metric. $\mathcal R$ is the curvature scalar and $\kappa^2=8\pi G$ is the coupling constant derived from the Newton's gravitational constant. The corresponding field equation is
\bea
\label{feq}
\mathcal R_{\mu\nu} \,-\,\half\, g_{\mu\nu}\mathcal R\,=\,0\ .
\eea
We now make the following {\it {three}} gauge choices~\cite{SS, BCL}
\bea \label{lcg}
g_{--}\,=\,g_{-i}\,=\,0\quad ,\; i=1,2\ .
\eea
\ndt These choices are motivated by the fact that $\eta_{--}\!=\eta_{-i}\!=0$. The metric is parametrized as 
\bea
\label{gc}
\begin{split}
g_{+-}\,&=\,-\,e^\phi\ , \\
g_{i\,j}\,&=\,e^\psi\,\gamma_{ij}\ .
\end{split}
\eea
$\phi, \psi$ are real parameters and $\gamma^{ij}$ is a two-by-two real, symmetric unimodular matrix. Field equations that do not involve time derivatives $(\del_+)$ are constraint relations as opposed to equations of motion, which explicitly contain time derivatives.
\vskip 0.1cm
\ndt The $\mu\!=\!\nu\!=\!-\;\;$equation from (\ref {feq}) is a constraint relation and yields 
\bea \label{CE1}
2\,\del_-\phi\,\del_-\psi\,-\,2\,\del^2_-\psi\,-\,(\del_-\psi)^2\,+\, \half \del_-\gamma^{ij}\,\del_-\gamma_{ij}\,=\,0\ .
\eea
This may be solved using our {\it {fourth}} gauge choice,
\bea 
\phi\,=\,\frac{\psi}{2}.
\eea 
From \eqref{CE1} 
\bea 
\label{psio}
\psi\,=\,\frac{1}{4}\,\frac{1}{\del^2_-}\,(\del_-\gamma^{ij}\,\del_-\gamma_{ij})\ .
\eea 
Other constraint relations are used to eliminate more (unphysical) metric components. For example, $\mu=i$, $\nu=\!-\;$ in (\ref {feq}) tells us that 
\bea \label{CE2}
g^{-i}&=&\mathrm{e}^{-\,\phi}\,\frac{1}{\partial_-}\bigg[\,\gamma^{ij}\,\mathrm{e}^{\phi\,-\,2\,\psi}\,\frac{1}{\partial_-}\,{\Big \{}\,\,\mathrm{e}^{\psi}\,{\Big (}\,\frac{1}{2}\,\partial_-\,\gamma^{kl}\,\partial_j\,\gamma_{kl}\,-\,\,\partial_-\,\partial_j\,\phi \nn\\
&&\,-\,\partial_-\,\partial_j\,\psi\,+\,\partial_j\phi\,\partial_-\,\psi\,\,{\Big )}\,+\,\,\partial_l\,{\Big (}\,\mathrm{e}^{\psi}\,\gamma^{kl}\,\partial_-\,\gamma_{jk}\,{\Big )}\,{\Big \}}\,\bigg]\ . 
\eea
\ndt The Einstein-Hilbert action now reads
\bea
\label{aaction}
S\,&=&\frac{1}{2\kappa^2}\int d^{4}x \; e^{\psi}\left(2\,\del_{+}\del_{-}\phi\, +\, \del_+\del_-\psi - \half\,\del_{+}\gamma^{ij}\del_{-}\gamma_{ij}\right) \nonumber \\
&&-e^{\phi}\gamma^{ij}\left(\del_{i}\del_{j}\phi + \half \del_{i}\phi\del_{j}\phi - \del_{i}\phi\del_{j}\psi - \frac{1}{4}\del_{i}\gamma^{kl}\del_{j}\gamma_{kl} + \half \del_{i}\gamma^{kl}\del_{k}\gamma_{jl}\right) \nn \\
&&- \half e^{\phi - 2\psi}\gamma^{ij}\frac{1}{\del_{-}}R_{i}\frac{1}{\del_{-}}R_{j}\ ,
\eea
\ndt where 
\bea 
R_{i}\,\equiv\, e^{\psi}\left(\half \del_-\gamma^{jk}\del_{i}\gamma_{jk}-\del_-\del_i\phi - \del_-\del_i\psi + \del_i\phi\del_-\psi\right)+\del_k(e^\psi\,\gamma^{jk}\del_-\gamma_{ij})\ . \nn
\eea
This closed form expression~\cite{BCL} describes gravitation, in light-cone gauge, purely in terms of its physical degrees of freedom.

\subsection{Perturbative expansion}
\vskip .3cm
\ndt We now perform a perturbative expansion of the closed form result in (\ref {aaction}). The order $\kappa^2$ result was first presented in~\cite{BCL} while the $\kappa^3$ vertices were derived in~\cite{SA}. We parameterize the matrix $\gamma_{ij}$ as
\bea 
\gamma_{ij}\,=\,(e^H)_{ij}\ ,
\eea
where $H$ is a traceless matrix since $det\,(\,\gamma_{ij})=1$. We choose
\bea \label{matrixH}
H\,=\,\begin{pmatrix} h_{11} & h_{12} \\ h_{12} & -h_{11} \end{pmatrix}\,; \quad h\,=\,\frac{(h_{11}+i\,h_{12})}{\sqrt{2}}\,, \,\,\,\,\,\,\,\,\,\,\,\,\, \bar{h}\,=\,\frac{(h_{11}-i\,h_{12})}{\sqrt{2}}\ ,
\eea
\ndt From (\ref {psio})
\bea 
\psi\,=\,-\frac{1}{\del^2_-}(\del_-h\,\del_-\bar h)\,+\, \mathcal{O} (h^4)\ ,
\eea
\ndt and we rescale all the fields
\bea 
\label{newfield}
h\,\rightarrow \, \frac{h}{\kappa}\ .
\eea
\ndt The Lagrangian (density) at lowest order now reads
\bea 
\label{kinetic}
\mathcal{L}_2\,=\, \half \,\bar h\,\Box\, h\ ,
\eea
\ndt with the d'Alembertian $\Box=2(\,\partial{\bar \partial}-\partial_+\parm\,)$. At order $\kappa$, we have
\bea 
\label{cubic}
\mathcal{L}\,=\, 2\,\kappa\, \bar{h}\, \parm^2\left[-\,h\,\frac{\bar{\del}^2}{\parm^2}h\,+\,\frac{\bar{\del}}{\parm}h\,\frac{\bar{\del}}{\parm}h\right] \,+\, \text{complex conjugate}\ .
\eea
\ndt At the next order, time derivatives need to be removed using a suitable field redefinition and the resulting quartic Lagrangian was presented in~\cite{BCL, ABHS}. After some simplifications, the Hamiltonian, to order $\kappa$,  corresponding to the Lagrangians above, may be written as
\bea 
\label{gqf}
\mathcal H\,=\,\int d^3 x\;\; \mathcal D\bar h\,\,\bar{\mathcal D}h\;\;,
\eea
where 
\bea \label{coderivative}
\mathcal D \bar h\,=\,\del\bar h \,+\,2\kappa\, \frac{1}{\del_-^2}\,\,\big( \frac{\bar\del}{\del_-} h\,\del_-^3 \bar h\,-\, h\,\del_-^2 \bar\del \bar h\big)\ ,
\eea
and $\bar{\mathcal D}h$ is the complex conjugate of $\D \bar h$.
\vskip 0.3cm

\subsection{Residual reparametrization invariance: order $\kappa^0$}

\vskip 0.3cm

\ndt To complete our description of gravity in light-cone gauge, we must examine the effect of residual reparametrizations. To lowest order in $\kappa$, these take the form
\be 
x \rightarrow x\ +\ \xi(\bar{x})\ , \qquad \bar{x} \rightarrow \bar{x}\ +\ \bar{\xi}(x)\ ,
\ee
\ndt By examining how the metric transforms, we find that the field transforms aa follows
\be 
\label {variation}
\delta h\ =\ \frac{1}{2 \kappa} \del \xi \ +\ \xi \bar{\del} h \ +\ \bar{\xi} \del h\ ,
\ee
where $\xi$ satisfies

\be \label{constraints}
\del_- \xi \ =\  0\ , \qquad \bar{\del}\ \xi \ =\ 0\ .
\ee
\ndt To order $\kappa^{-1}$ we have
\be \label{extra constraints}
\del_- (\delta h) \ =\ 0\ , \qquad \bar{\del}  (\delta h) \ =\ 0\ .
\ee
\vskip 0.3cm
\ndt The variation of the Hamiltonian to order $\kappa^0$ is
\be 
\label{zero order}
\delta \mathcal{H}^{(\kappa^0)} \ =\ \delta\ ( \del\bar h\,\bar{\del}h ) +\  2\kappa\ \delta^{\kappa^{-1}}\ \left\{\ \bar{h} \ {\del_-}^2 \left( h \frac{\bar{\del}^2}{{\del_-}^2} h\ -\ \frac{\bar{\del}}{\del_-}h \frac{\bar{\del}}{\del_-}h\ \right) \ +\;{\mbox {c.c.}}\; \right\}\ ,
\ee
\ndt with the first term in (\ref{zero order}) yielding
\be 
\label{linear}
\begin{split}
-\ \del \xi\ \bar{h}\ \bar{\del}^2 h\ -\ \bar{\del} \bar{\xi}\ h\ \del^2 \bar{h}\ .
\end{split}
\ee
\ndt The variation of the second term in (\ref{zero order}) and its complex conjugate exactly cancel the terms above, proving that
\be
\delta \mathcal{H}^{(\kappa^0)}\ =\ 0\ .
\ee
\ndt The Hamiltonian in (\ref {gqf}) is thus invariant under the following transformations
\be
\label{kappa corr a} 
\delta h\ =\ \frac{1}{2 \kappa} \del \xi \ +\ \xi \bar{\del} h \ +\ \bar{\xi} \del h\ ,
\ee
\be
\label{kappa corr b} 
\delta \bar{h}\ =\ \frac{1}{2 \kappa} \bar{\del} \bar{\xi} \ +\ \xi \bar{\del} \bar{h} \ +\ \bar{\xi} \del \bar{h}\ .
\ee
\ndt We see now that the derivative introduced in (\ref {coderivative}) transforms `covariantly'. That is
\bea
\delta (\bD h)\,=\,(\xi \bar{\del} +\ \bar{\xi} \del\,)\,\bD\,h\ ,
\eea
at this order, in keeping with the analysis of Yang-Mills theory~\cite{ABM}. 

\vskip 0.1cm

\section{The Hamiltonian, to order $\kappa^2$}
\vskip 0.3cm

\ndt Moving to order $\kappa^2$, the Hamiltonian is~\cite{ABHS}
\bea 
\label{hamiltonian}
\mathcal H&=&\,\del \bar h\,\bar \del h \,-2\,\kappa\, \bar{h}\, \parm^2\left\{-\,h\,\frac{\bar{\del}^2}{\parm^2}\ h\,+\,\frac{\bar{\del}}{\parm}h\,\frac{\bar{\del}}{\parm}h\right\}\ - 2\,\kappa\, h\, \parm^2\left\{-\,\bar h\,\frac{\del^2}{\parm^2}\ \bar h\,+\,\frac{\del}{\parm}\bar h\,\frac{\del}{\parm}\bar h\right\} \nn \\ [.2cm]
&&-\,4\kappa^2\,{\Bigg\{}-2\,\frac{1}{\del_-^2}\bigg(\frac{\bar \del}{\del_-}\,h\,\del_-^3\bar h\,-\,h\,\del_-^2\bar\del\bar h\bigg)\frac{1}{\del_-^2}\bigg(\frac{\del}{\del_-}\,\bar h\,\del_-^3 h\,-\,\bar h\,\del_-^2 \del h\bigg)\nn\\[.2cm]
&&+\frac{1}{\del_-^2}(\bar\del h\,\del_-^2\bar h\,-\,\del_-h\,\del_-\bar \del\bar h)\,\frac{1}{\del_-^2}\,(\del \bar h\,\del_-^2 h\,-\,\del_- \bar h\,\del_- \del h)\,-\,3\,\frac{1}{\del_-}(\bar \del h\,\del_-\bar h)\,\frac{1}{\del_-}\,(\del_-h\,\del \bar h) \nn\\[.2cm]
&&+\,\frac{1}{\del_-}(\bar \del h\,\del_-\bar h\,-\,\del_-h\,\bar \del\bar h)\,\frac{1}{\del_-}\,(\del \bar h\,\del_- h\,-\,\del_- \bar h\, \del h)\,+\, 3\,\frac{1}{\del_-}\,(\del_- h\,\del_-\bar h)\,\frac{1}{\del_-}\,(\bar \del h\,\del \bar h)\nn\\[.2cm]
&&+\,\bigg[\frac{1}{\del_-^2}(\del_-h\,\del_-\bar h)\,-\,h\,\bar h\bigg](\bar \del h\,\del \bar h\,+\,\del h\, \bar\del\,\bar h\,-\,\del_-h\,\frac{\del\,\bar \del}{\del_-}\,\bar h\,- \,\del_-\bar h\,\frac{\del\,\bar \del}{\del_-}\, h){\Bigg\}}\ .
\eea

\vskip 0.7cm

\subsection{Residual reparametrization invariance: order $\kappa$}
\label{rriq}
\vskip 0.3cm

\ndt Once quartic interaction vertices are included, the resulting Hamiltonian (\ref {hamiltonian}) is no longer invariant under the infinitesimal symmetry transformations introduced earlier. To see this, consider the relevant contributions from the cubic and quartic vertices. 
\be
\label{cubquart}
\delta \mathcal{H}_{c,q}^{(\kappa)}\ =\ \delta^{\kappa^0} (\text{cubic terms})\ +\ \delta^{\kappa^{-1}} (\text{quartic terms})\ .
\ee
We present details of this calculation in appendix A. We find that the net contribution, from the cubic and quartic vertices, at order $\kappa$ is
\be  
\label{quartic}
\delta \mathcal{H}_{c,q}^{(\kappa)}\ =\ \left(+2\kappa\ \bar{\del}\bar{\xi}\ h\ \del h\ \bar{\del} \bar{h}\ -\  2\kappa h\ \bar{\del}\bar{\xi}\ \del_- \bar{h}\ \frac{\del \bar{\del}}{\del_-}h\ \right)\,+\,{\mbox {c.c.}}
\ee
\ndt It is therefore clear that the existing transformations in (\ref {variation}) do not leave the Hamiltonian invariant. In order to render it invariant, we are forced to introduce new terms at order $\kappa$, to the r.h.s of (\ref {variation}). These new contributions, when substituted in the kinetic term in (\ref {hamiltonian}), are clearly at the same order as those in (\ref{quartic}). We find
\be\label{kappa corr 1} 
\delta h\ =\ \frac{1}{2 \kappa} \del \xi \ +\ \xi \bar{\del} h \ +\ \bar{\xi} \del h\ \ -\ \kappa\ \bar{\del} \bar{\xi}\ h h\ +\ 2\kappa\ \del \xi\ \frac{1}{\del_-}(\bar{h}\ \del_- h)\ ,
\ee
\ndt
and
\be\label{kappa corr 2} 
\delta \bar{h}\ =\ \frac{1}{2 \kappa} \bar{\del} \bar{\xi} \ +\ \xi \bar{\del} \bar{h} \ +\ \bar{\xi} \del \bar{h}\ \ -\ \kappa\ \del \xi\ \bar{h} \bar{h}\ +\ 2\kappa\ \bar{\del}  \bar{\xi}\ \frac{1}{\del_-}(h\ \del_- \bar{h})\ .
\ee
\vskip 0.2cm
\ndt The variation $\,\delta^{\kappa} (\del\bar{h}\, \bar{\del} h)\,$ cancels exactly against the terms in (\ref{quartic}), confirming that
\be
\delta \mathcal{H}^{(\kappa)}\ =\ 0\ ,
\ee
and proving invariance of the light-cone Hamiltonian, to order $\kappa^2$, under the residual reparametrizations (\ref{kappa corr 1}) and  (\ref{kappa corr 2}). 
\vskip 0.3cm
\ndt The transformations to order $\kappa^0$ were used to identify counter terms for gravity to the appropriate order. In particular~\cite{BBK}
\be
[\delta_1(\xi_1),\delta_2(\xi_2)]\,h=\delta_{12}(\xi_{12})\,h\ ,
\ee
where the resulting parameter is
\be
\label{parameter} 
\xi_{12} = \bar \xi_2 \partial \xi_1 - \bar \xi_1 \partial \xi_2.
\ee
This parameter does not satisfy (\ref{constraints}) because we discard a determinant of $\partial_-$ in the functional integral when we integrate out the unphysical degrees of freedom. We have to restore these to have a finite residual reparametrization. However, the infinitesimal symmetries are sufficient to constrain the Hamiltonian~\cite{BBK}.
\vskip 0.3cm
\ndt We can now check closure of the full transformation to order $\kappa$ and indeed, it still closes to the same parameter (\ref{parameter}), showing that it is indeed a residual reparametrization symmetry. We thus see that these are the first few terms in an infinite series which represents the entire infinitesimal residual reparametrization symmetry. The fact that it constrains possible terms in the Hamiltonian shows that it works just like a finite symmetry in this respect, and should be important for constraining loop expressions. The price we pay for not being able to integrate the symmetry is that we cannot use it to classify the invariants. We intend to return to this point, and examine this symmetry at null-like infinity to establish connections with~\cite{BMS}.

\vskip 1cm

\subsection{Quadratic form structure}
\label{qfs}
\vskip 0.1cm

\ndt In this subsection, we demonstrate that the Hamiltonian, to order $\kappa^2$, is a quadratic form. That is
\bea 
\mathcal H\,=\,\int d^3 x\;\; \mathcal D\bar h\,\,\bar{\mathcal D}h\;\;.
\eea
\ndt From each line of the Hamiltonian  in (\ref {hamiltonian}), we will compute contributions to $\D \bar h$ (we already know $\D \bar h$ to order $\kappa$). The product of the order $\kappa$ terms, $\D \bar h\,(\kappa)\;\bD h\,(\kappa)$, yields one-half of the second line in (\ref {hamiltonian}). We need to show then that half of the second line in (\ref {hamiltonian}) and all the remaining terms, of order $\kappa^2$ may be rewritten in the form
\bea
\D \bar h\,(\kappa^2)\,\bar\del h\,+\,\del \bar h\,\bD h\,(\kappa^2)\ .
\eea

\subsubsection*{from line 2}

\vskip 0.1cm
\ndt Contribution to $\mathcal D \bar h$

\bea
&&+2\,\kappa^2\,\fr{\parm}\{\parm^2\bar h\fr{\parm^3}(\parm^3 h\frac{\partial}{\parm}\bar h-\parm^2\partial h \bar h\,)\,\} \\
&&+2\,\kappa^2\,\fr{\parm}\{\,\frac{\partial}{\parm^4}(\bar h\parm^2 h)\parm^3 \bar h\}
\eea

\ndt The remaining terms (that cannot immediately be written in the form $X\bar\del h$ or $Y\del \bar h$) are
\bea
-2\,\kappa^2\,h\parm^2\bar h \,\frac{\bar\partial}{\parm^4}(\parm^2\partial h \bar h)+{\mbox {c.c.}}
\eea

\ndt The rest of this calculation is presented in appendix B and we simply state here, the result for $\mathcal D\bar h$. 
\vskip 0.1cm
\ndt At order $\kappa^2$, $\mathcal D \bar h$ reads

\bea
\label{covd}
&&+2\,\kappa^2\,\fr{\parm}\{\parm^2\bar h\fr{\parm^3}(\parm^3 h\frac{\partial}{\parm}\bar h-\parm^2\partial h \bar h\,)\,\}+2\,\kappa^2\,\fr{\parm}\{\,\frac{\partial}{\parm^4}(\bar h\parm^2 h)\parm^3 \bar h\} \nn \\
&&-2\,\kappa^2\,\parm^2\bar h\,\fr{\parm^4}(\parm^2 h\partial \bar h-2\parm \partial h \parm \bar h\,)+2\,\kappa^2\,\parm \bar h \fr{\parm^2}(\parm h \partial \bar h -2 \partial h \parm \bar h) \nn \\
&&+6\,\kappa^2\,\fr{\parm^2}(\parm h \parm \bar h)\,\partial \bar h-6\,\kappa^2\,\parm \bar h\,\fr{\parm^2}(\parm h \partial \bar h)-2\,\kappa^2\,\fr{\parm^2}(\parm h\parm \bar h)\partial \bar h \nn \\
&&+4\kappa^2\,h\bar h\,\partial \bar h+4\kappa^2\,\frac{\partial}{\parm}\{\parm \bar h\,(\fr{\parm^2}(\parm h\parm \bar h)-h \bar h)\,\}+2\kappa^2\,\parm^2\bar h\fr{\parm^4}(\parm^2\partial h \bar h) \nn \\
&&-2\kappa^2\,\parm\{\parm \bar h\fr{\parm^2}(\bar h \partial h)\}-2\kappa^2\,\partial\{\bar h\fr{\parm^2}(\parm \bar h \parm h)\}-2\kappa^2\,\parm^2\bar h\fr{\parm^3}(\parm\partial h \bar h) \nn \\
&&+2\kappa^2\parm\partial\{\bar h \fr{\parm^3}(h \parm^2 \bar h)\}+2\kappa^2\partial\{\parm\bar h\fr{\parm^3}(\bar h \parm^2 h)\}+2\kappa^2\parm^2\{\bar h\fr{\parm^3}(\parm \bar h \partial h)\}\ ,
\eea
\vskip 0.1cm
\ndt confirming that the light-cone Hamiltonian for pure gravity, in $d=4$, may be expressed as a quadratic form up to order $\kappa^2$.
\vskip 0.3cm
\ndt Like at order $\kappa$, one might expect $\mathcal D\bar h$ in (\ref {covd}) to transform covariantly. Unfortunately, at this order, this does not happen - the derivative does not transform like the field. Explicit variation of (\ref {covd}) yields
\bea
\label{vard1}
\delta\,(\,\D \bar h\,)^\kappa=&&+\kappa\,\del\xi\,\del\{\,\del_-\bar h \fr{\parm}\bar h\,\} \nn \\
&&+\,2\kappa\,\bar\del\bar\xi\,h\,\del\,\bar h\,+\,\kappa\,\bar\del\,\bar\xi\,\del\,\del_-\,\bar h\,\fr{\parm}\,h\,-\,\kappa\,\bar\del\,\bar\xi\,\fr{\parm}\,\{\,\del_-\,\del\,\bar h\,h\,\}\ ,
\eea
using which it is easy to verify that
\bea
\label{2two}
\delta \mathcal{H}^\kappa\,=\,\int d^3 x\ {[\,\delta (\D \bar h)\,\bD h\ +\ \D \bar h\, \delta(\bD h)\,]}^\kappa\,=\,0\ .
\eea
In the next section we explain, on general grounds, why the transformation property in (\ref {vard1}) is not unexpected.

\vskip 1cm
\subsection{Transformation properties of $\mathcal D\bar h$}
\vskip 0.3cm

\ndt Based on helicity considerations, (\ref{kappa corr 1}) and dimensional analysis, we start with the following general ansatz for $\delta({\bD h})$ 
\be 
\label{ansatz1}
\delta (\bD h)=0+\,(\,\xi\bar\del\,+\,\bar{\xi} \del\,)\,\bD h\,-\,\kappa\,\bar{\del} \bar{\xi}\ \sum_i\,\alpha_i\,\hat{A_i}\ (\hat{B_i}h\ \hat{C_i} h)\,+\,2\kappa\,\del\xi\,\sum_j\,\beta_j\, {\hat{P_j}}({\hat{Q_j}}\bar h\,{\hat{R_j}} h)\ ,
\ee
\ndt and for its complex conjugate
\be 
\label{ansatz2}
\delta (\D \bar{h})\ =0\,+\,(\,\xi\bar\del\,+\,\bar{\xi}\del\,)\,\D\bar h\,-\,\kappa\,\del\xi\,\sum_i\,\alpha_i\,\bar{\hat {A_i}}\,(\bar {\hat {B_i}}\bar h\,\bar {\hat {C_i}}\bar h\,)\,+\,\ 2\kappa\ \bar{\del}  \bar{\xi}\ \sum_j\,\beta_j\,\bar{\hat{P_j}}(\bar{\hat{Q_j}}h\bar{\hat{R_j}}\bar{h})\ .
\ee
\vskip -0.3cm
\ndt The $\hat A_i,\ldots$ are operators to be determined later while $\alpha_i$ and $\beta_j$ are constants. Note that this ansatz transforms covariantly if we choose the following single set of operators
\bea
\label{choose}
&&\alpha=1\,,\;\hat{A}\ =\ \bar \partial\ ,\ \hat{B}\ =\ \hat{C}\ =\ 1\ , \nn \\
&&\beta=1\,,\;\hat{P}\ =\ \frac{1}{\parm}\ , \ \hat{Q}\ =\ 1\ , \ \hat{R}\ =\ \parm \bar \partial\ .
\eea
Since the Hamiltonian is invariant under (\ref{kappa corr 1}) and (\ref{kappa corr 2}), we have
\bea
\label{1one}
\delta \mathcal{H}= 0\ \implies \int d^3 x\ [\,\delta (\D \bar h)\,\bD h\ +\ \D \bar h\, \delta(\bD h)\,]=0\ .
\eea
\ndt
Let us first verify (\ref {1one}) at order $\kappa^0$
\bea
\delta \mathcal{H}=\!\!\!\!\!\!\!\!&&\int d^3 x\ [\,(\delta (\D \bar h))^{\kappa^0}\bar{\partial} h\ + \partial \bar h (\delta(\bD h))^{\kappa^0}\,]\ ,\\
=\!\!\!\!\!\!\!\!&& \int d^3 x\  [\,\bar \xi \partial^2 \bar h \bar \partial h \ +\ \partial \bar h \bar \xi \partial \bar \partial h\,]\ .
\eea
\ndt Integrating a $\partial$ from the $\bar h$ in the first term yields $(\delta \mathcal{H})^{\kappa^0}\ =\ 0$.
\vskip 0.3cm
\ndt
At order $\mathcal{O}(\kappa)$, we have
\bea
(\delta \mathcal{H})^\kappa &=&\int d^3 x\ [\,(\delta (\D \bar h))^{\kappa}\bar{\partial} h+(\delta (\D \bar h))^{\kappa^0}(\bD h)^{\kappa}+(\D \bar h)^\kappa (\delta(\bD h))^{\kappa^0}+\partial \bar h (\delta(\bD h))^\kappa\,]\ , \nn\\
&=&\kappa \int d^3 x\ \{\, [\, \bar \xi \partial (\D \bar h)^\kappa +\,\ 2\bar{\del}  \bar{\xi}\ \sum_j\,\beta_j\,\bar{\hat{P_j}}(\bar{\hat{Q_j}}h\bar{\hat{R_j}}\bar{h})\,]\bar \partial h\ +\ \bar \xi \partial^2 \bar h (\bD h)^\kappa\ \label{X1} \\
&&\quad +\ [\bar \xi \partial (\bD h)^\kappa-\,\bar{\del} \bar{\xi}\ \sum_i\,\alpha_i\,\hat{A_i}\ (\hat{B_i}h\ \hat{C_i} h)]\ \partial \bar h\ +\ (\D \bar h)^\kappa \bar \xi \partial \bar \partial h  \,\} \label{Y1}\ .
\eea
We integrate a $\partial$ from $\bar h$ in the last term of (\ref{X1}) to cancel it against the first term in (\ref{Y1}). We then cancel last term of (\ref{Y1}) against the first term of (\ref{X1}) by integrating a $\partial$. Note that the exact form of $\mathcal D\bar h$, at order $\kappa$, is irrelevant to this analysis. We are left with
\bea 
\label{clever}
(\delta \mathcal{H})^\kappa=\kappa\int d^3 x\ [\,2\bar{\del}  \bar{\xi}\ \sum_j\,\beta_j\,\bar{\hat{P_j}}(\bar{\hat{Q_j}}h\bar{\hat{R_j}}\bar{h}) \bar \partial h\  -\,\bar{\del} \bar{\xi}\ \sum_i\,\alpha_i\,\hat{A_i}\ (\hat{B_i}h\ \hat{C_i} h)\,\partial \bar h ]\ .
\eea
\ndt Substituting (\ref {choose}) into (\ref {clever}) , we find
\bea
(\delta \mathcal{H})^\kappa= +2 \,\kappa\,\int d^3 x\;\, \bar \partial \bar \xi\, \frac{1}{\parm} (\,\parm h\,\partial \bar h\,)\,\bar \partial h\,+\,{\mbox {c.c.}}\,\neq\, 0
\eea
Thus the Hamiltonian for gravity is a quadratic form but it is not the ``square'' of a covariant derivative. Instead, if we choose the operators $\hat A_i\ldots$ and constants appropriate to (\ref {vard1}) then (\ref {clever}) yields
\bea
(\delta \mathcal{H})^\kappa\,=\,0\ .
\eea

\ndt This is a clear point of contrast from Yang-Mills theory where both the pure and maximally supersymmetric theories are described by quadratic form structures composed of covariant derivatives~\cite{ABM}. From the MHV literature~\cite{MHV}, we know that all tree-level scattering amplitudes in Yang-Mills theory may be expressed entirely in terms of the ``square" or ``angular" brackets. In gravity, the cubic amplitude does indeed have the same property but the quartic and higher vertices involve a mixture of both brackets.The derivatives we have introduced in the quadratic form for gravity do not transform covariantly beyond order $\kappa$ and this is very likely, another way of stating what the amplitude structures have already taught us.

\vskip 0.7cm

\section{Quadratic forms in $\mathcal N=8$ supergravity}

\vskip 0.1cm

\ndt In this section, we review aspects of $\mathcal N=8$ supergravity essential to our understanding of its light-cone Hamiltonian and the corresponding quadratic form structure. The physical degrees of freedom in $\mathcal N=8$ supergravity are all described by a single superfield~\cite{BLN} (and its conjugate) with Grassmann variables $\theta^m$ ($m \ldots$ are $SU(8)$ indices)
\bea\label{superfield}
\begin{split}
\phi\,(\,y\,)\,=&\,\frac{1}{\parm^2}\,h\,(y)\,+\,i\,\theta^m\,\frac{1}{\parm^2}\,{\bar \psi}_m\,(y)\,-\,\frac{i}{2}\,\theta^m\,\theta^n\,\frac{1}{\parm}\,{\bar A}_{mn}\,(y)\ , \\
\;&+\,\frac{1}{3!}\,\theta^m\,\theta^n\,\theta^p\,\frac{1}{\parm}\,{\bar \chi}_{mnp}\,(y)\,-\,\frac{1}{4!}\,\theta^m\,\theta^n\,\theta^p\,\theta^q\,{\bar C}_{mnpq}\,(y)\ , \\
\;&+\,\frac{i}{5!}\,\theta^m\,\theta^n\,\theta^p\,\theta^q\,\theta^r\,\epsilon_{mnpqrstu}\,\chi^{stu}\,(y)\ ,\\
\;&-\,\frac{i}{6!}\,\theta^m\,\theta^n\,\theta^p\,\theta^q\,\theta^r\,\theta^s\,\epsilon_{mnpqrstu}\,\parm\,A^{tu}\,(y)\ ,\\
\,&-\,\frac{1}{7!}\,\theta^m\,\theta^n\,\theta^p\,\theta^q\,\theta^r\,\theta^s\,\theta^t\,\epsilon_{mnpqrstu}\,\parm\,\psi^u\,(y)\ ,\\
\,&+\,\frac{4}{8!}\,\theta^m\,\theta^n\,\theta^p\,\theta^q\,\theta^r\,\theta^s\,\theta^t\,\theta^u\,\epsilon_{mnpqrstu}\,\parm^2\,{\bar h}\,(y)\ ,
\end{split}
\eea
the fields being the graviton, the gravitinos, the gauge fields, the gauginos and the scalar fields. All fields are local in
\bea
y~=~\,(\,x,\,{\bar x},\,{x^+},\,y^-_{}\equiv {x^-}-\,\frac{i}{\sqrt 2}\,{\theta_{}^m}\,{{\bar \theta}^{}_m}\,)\ .
\eea
The superfields satisfy
\be
d^m\,\phi\,(\,y\,)\,=\,0\;\; ;\qquad {\bar d}_n\,{\bar \phi}\,(\,y\,)\,=\,0\ ,
\ee
\noindent where
\bea
d^{\,m}\,=\,-\,\frac{\partial}{\partial\,{\bar \theta}_m}\,+\,\frac{i}{\sqrt 2}\,\theta^m\,\parm\;\; ;\qquad{\bar d}_n\,=\,\frac{\partial}{\partial\,\theta^n}\,-\,\frac{i}{\sqrt 2}\,{\bar \theta}_n\,\parm\ ,
\eea
\ndt are chiral derivatives. The fields also satisfy the inside-out constraint
\bea \label{inside-out}
\label{io}
\,{\phi}\,=\,\frac{1}{4}\,\frac{{(d\,)}^8}{{\parm}^4}\,{\bar \phi}\ ,
\eea
\noindent where ${(d\,)}^8\,=\,d^1\,d^2\,\ldots\,d^8$; a unique feature of maximally supersymmetric theories. At $x^+=0$,  the kinematic generators of the superPoincar\'e algebra are the three momenta, the transverse space rotation and the rotations, $j^+$, its conjugate and $ j^{+-}$~\cite{BBB}. The dynamical generators include the light-cone Hamiltonian
\be
p^-_{}~=~i\frac{\partial\bar\partial}{\partial_-}\ ,
\ee
and the dynamical boosts~\cite{ABKR}. Supersymmetry generators also come in two varieties, the kinematical
\bea
q^m_{\,+}\,=\,-\,\frac{\partial}{\partial\,{\bar \theta}_m}\,-\,\frac{i}{\sqrt 2}\,\theta^m\,\parm ;\qquad {\bar q}_{\,+\,n}=\;\;\;\frac{\partial}{\partial\,\theta^n}\,+\,\frac{i}{\sqrt 2}\,{\bar \theta}_n\,\parm\ ,
\eea
\noindent and the dynamical
\bea
\label{dynsus}
\begin{split}
Q_-^{\,m}\,\equiv\,&i\,[\,\bar j^-\,,\,q^m_{\,+}\,]\,=-\,\frac{\bar \partial}{\parm}\,q^m_{\,+}\,  , \\
{\bar Q}_{-\,n}\,\equiv\,&i\,[\,j^-\,,\,{\bar q}_{\,+\,n}\,]\,=-\,\frac{\partial}{\parm}\,{\bar q}_{\,+\,n}\, .
\end{split}
\eea

\subsection{The action to order $\kappa$}
\vskip 0.3cm

\noindent In this light-cone superspace formalism, the ${\mathcal N}=8$ supergravity action to order $\kappa$ reads

\be
\label{n=8}
\beta\,\int\;d^4x\,\int d^8\theta\,d^8 \bar \theta\,{\cal L}\ ,
\ee
where $\beta\,=\,-\,\frac{1}{64}$ and
\bea
\label{one}
{\cal L}&=&-\bar\phi\,\frac{\Box}{{\partial_-}^{4}}\,\phi\,-\,2\,\kappa\,(\,\frac{1}{{\parm}^2}\;{\nbar \phi}\;\;{\bar \partial}\,{\phi}\;{\bar \partial}\,{\phi}+\,\frac{1}{{\parm}^2}\;\phi\,\partial\,{\nbar \phi}\,\partial\,{\nbar \phi})\ .
\eea

\noindent The dynamical supersymmetry generator in (\ref {dynsus}) involves a new contribution at order $\kappa$,

\be\label{Q}
\bar Q_m{}^{(\kappa)} \phi= - \frac{1}{\parm}(\bar \partial \bar q_m \phi {\parm}^2 \phi - \parm \bar q_m \phi \parm \bar \partial \phi).
\ee
(where the $+$ index on $q_+$ is not shown). The complex conjugate yields $Q^m{}^{(\kappa)} \bar \phi$ and the inside-out constraint determines $Q^m{}^{(\kappa)} \phi$ and $\bar Q_m{}^{(\kappa)} \bar \phi$. The anticommutator of the dynamical supersymmetry generators is the light-cone Hamiltonian.

\vskip 0.3cm
\subsection{Hamiltonian written as a quadratic form}
\vskip 0.5cm
\noindent At lowest order~\cite{ABHS}
\bea
\label{claim}
{\cal H}~=~\frac{1}{4\,\sqrt{2}}\,(\,{\mathcal W}_{\,m}\,,\,{\mathcal W}_{\,m}\,)\ ,
\eea
with
\bea
{\mathcal W}_{\,m}\,=\,{\bar Q}_{-\,m}\,\phi\ ,
\eea
and
\be
\label{inner}
(\,\phi\,,\,\xi\,)~\equiv-~2i\int d^4\!x\, d^8\theta\,d^8\,{\bar\theta}\;{\bar\phi}\,\frac{1}{{\parm}^3}\xi\ .
\ee

\noindent Note that (\ref {claim}) is unrelated to the fact that the Hamiltonian is the anticommutator of two supersymmetries. To see this, start with (\ref {claim})
\bea
\begin{split}
{\cal H}^0\,&=\,\frac{1}{4\,\sqrt{2}}\,(\,{\mathcal W}^0_m\,,\,{\mathcal W}^0_m\,)\ , \\
&=-\,\frac{2i}{4\,\sqrt{2}}\,\int d^4\!x\, d^8\theta\,d^8\,{\bar\theta}\;\;Q_-^{\,m}\,{\bar \phi}\,\frac{1}{{\parm}^3}\,{\bar Q}_{-\,m}\,\phi\ ,
\end{split}
\eea
\noindent and rewrite it as two terms
\bea
\label{unik}
{\cal H}^0\,=-\,\frac{i}{4\,\sqrt{2}}\,\,\int d^4\!x\, d^8\theta\,d^8\,{\bar\theta}\;\;{\Big (}\,Q_-^{\,m}\,{\bar \phi}\,\frac{1}{{\parm}^3}\,{\bar Q}_{-\,m}\,\phi\,+\,Q_-^{\,m}\,{\bar \phi}\,\frac{1}{{\parm}^3}\,{\bar Q}_{-\,m}\,\phi\,{\Big )}\ .
\eea
\noindent In maximally supersymmetric theories alone, we have constrained superfields obeying (\ref{io}). We use this in the second term in equation (\ref {unik}) to obtain
\bea
{\cal H}^0\,=-\,\frac{i}{4\,\sqrt{2}}\,\int d^4\!x\, d^8\theta\,d^8\,{\bar\theta}\;\;{\Big (}\,Q_-^{\,m}\,{\bar \phi}\,\frac{1}{{\parm}^3}\,{\bar Q}_{-\,m}\,\phi\,+\,\frac{1}{{\parm}^4}\,Q_-^{\,m}\,\phi\,\parm\,{\bar Q}_{-\,m}\,{\bar \phi}\,{\Big )}\ .
\eea
\noindent Using (\ref {dynsus}) yields
\bea
{\cal H}^0\,=-\,\frac{i}{4\,\sqrt{2}}\,\,\int d^4\!x\, d^8\theta\,d^8\,{\bar\theta}\;\;{\Big (}\,\frac{\bar \partial}{\parm}\,q^m_{\,+}\,{\bar \phi}\,\frac{\partial}{{\parm}^4}\,{\bar q}_{\,+\,m}\,\phi\,+\,\frac{\bar \partial}{{\parm}^5}\,q^m_{\,+}\,\phi\,\partial\,{\bar q}_{\,+\,m}\,{\bar \phi}\,{\Big )}\ .
\eea
\noindent which we integration by parts 
\bea
{\cal H}^0\,=-\,\frac{i}{4\,\sqrt{2}}\,\int d^4\!x\, d^8\theta\,d^8\,{\bar\theta}\,\frac{\partial\,{\bar \partial}}{{\parm}^5}\,{\bar \phi}\,\{\,q^m_{\,+}\,,\,{\bar q}_{\,+\,m}\,\}\,\phi\ .
\eea
\noindent Since $\{\,q^m_{\,+}\,,\,{\bar q}_{\,+\,m}\,\}\,\phi\,=-\,i\,8\,{\sqrt 2}\,\parm\,\phi$, we have
\bea
{\cal H}^0\,=\,\int{d^4}x\,{d^8}\theta\,{d^8}{\bar \theta}\,
{\bar \phi}\,\frac{2\,\partial\bar\partial}{{\parm}^4}\,\phi\ ,
\eea
\noindent which is the expected kinetic term for the Hamiltonian. We refer the reader to~\cite{BKR} for details at order $\kappa^2$ and simply reproduce the relevant results here. At order $\kappa$ we have
\bea
\label{dublew}
{\mathcal W}_m\,=-\,\frac{\partial}{\parm} {\bar q}_{+\,m}\,\phi\,-\,\kappa\,\frac{1}{\parm}\,{\Big (}\,{\bar \partial}\,{\bar d}_m\,\phi\,{{\parm}^2}\,\phi\,-\,\parm\,{\bar d}_m\,\phi\,\parm\,{\bar \partial}\,\phi\,{\Big )}\,+\,{\cal O}(\kappa^2)\ ,
\eea
\bea
\label{dublewbar}
{\nbar {\mathcal W}}^m\,=-\,\frac{\bar \partial}{\parm}\,q_+^{\,m}\,{\bar \phi}\,-\,\kappa\,\frac{1}{\parm}\,{\Big (}\,\partial\,d^m\,{\bar \phi}\,{{\parm}^2}\,{\bar \phi}\,-\,\parm\,d^m\,{\bar \phi}\,\parm\,\partial\,{\bar \phi}\,{\Big )}\,+\,{\cal O}(\kappa^2)\ .
\eea
\noindent With these, we may directly compute the Hamiltonian as a quadratic form
\bea
\frac{1}{4\,\sqrt{2}}\,(\,{\mathcal W}\,,\,{\mathcal W}\,)\,=-\,\frac{2\,i}{4\,\sqrt{2}}\int d^4\!x\, d^8\theta\,d^8\,{\bar\theta}\;{\nbar {\mathcal W}}\,\frac{1}{{\parm}^3}\,{\mathcal W}\ .
\eea

\vskip 1cm

\section{Conclusions}

\noindent It is a somewhat surprising fact that the maximally supersymmetric and non-supersymmetric versions of both Yang-Mills theory and gravity (and only those) are quadratic forms. Note that this is over and above the residual reparametrization invariance  discussed earlier. What is puzzling is why the other supersymmetric versions (with supersymmetry but less than maximal supersymmetry) cannot be expressed in this form. It is possible that supersymmetric truncation~\cite{BT} 
\bea
\label{trunc1}
\int\,{d^4}x\,{d^8}\theta\,{d^8}{\bar \theta}\,{\cal L}\;=\;\frac{1}{16}\,\int\,{d^4}x\,{d^7}\theta\,{d^7}{\bar \theta}\,{\bar d}_4\,d^4\,{\cal L}\,\bigg|_{\theta^8\,=\,{\bar \theta}_8\,=0}\ ,
\eea
explains a portion of our results but it seems insufficient for a complete explanation. We are thus confident that this unique property signals behavior special to these theories. The place we hope to see this reflected is in loop-integrals. This is a macroscopic property and hence difficult to relate to some infinitesimal symmetry. We hope to return to this issue with a more refined mathematical analysis and understand its consequences for the quantum theories.
\vskip 0.3cm
\ndt The other loose end here is the infinitesimal residual reparametrisation invariance. For the analysis of this paper it was sufficient to consider the infinitesimal symmetry transformations, but it will be interesting to study this symmetry as an asymptotic symmetry of the theory and perhaps establish links with the work in~\cite{BMS}.

\vskip 1cm
\ndt {\it \bf {Acknowledgments}}
\vskip 0.1cm

\ndt NS acknowledges support from a DST Inspire fellowship. The work of MM is supported by the DST-Max Planck Partner Group on Cosmology and Gravity. SM acknowedges support from a CSIR NET fellowship. The work of SA is partially supported by a DST-SERB grant (EMR/2014/000687).

\vskip 1cm

\newpage

\renewcommand{\theequation}{A-\arabic{equation}}
\setcounter{equation}{0}  
\section*{Appendix {\bf {A}}}  
\vskip 0.5cm

\subsection*{Invariance of the Hamiltonian to order $\kappa$}
\vskip 0.5cm
We detail here, the computation relevant to subsection (\ref {rriq}). We start by varying the cubic terms.
\be 
\label{cubic2}
\begin{split}
&\delta^{\kappa^0}(\text{cubic terms})\  =\ 2\kappa (\bar{\xi}\del \bar{h}\ +\ \xi \bar{\del}\bar{h}){\del_-}^2 \left( h \frac{\bar{\del}^2}{{\del_-}^2} h\ -\ \frac{\bar{\del}}{\del_-}h \frac{\bar{\del}}{\del_-}h\ \right)\\\\
&\qquad +\ 2\kappa \bar{h}\ {\del_-}^2 \left( (\xi \bar{\del}h\ +\ \bar{\xi}\del h)\frac{\bar{\del}^2}{{\del_-}^2}h \ +\ h\frac{\bar{\del}^2}{{\del_-}^2}(\xi \bar{\del}h\ +\ \bar{\xi}\del h)\ -\ 2\ \frac{\bar{\del}}{\del_-} (\xi \bar{\del}h\ +\ \bar{\xi}\del h) \frac{\bar{\del}}{\del_-}h\right)\ ,\\\\
&\ =\ 2\kappa\ \bar{\xi}\ \del \bar{h}\ {\del_-}^2 \left( h \frac{\bar{\del}^2}{{\del_-}^2} h\ -\ \frac{\bar{\del}}{\del_-}h \frac{\bar{\del}}{\del_-}h\ \right)\\\\
&\qquad +\ 2\kappa \bar{h}\ {\del_-}^2 \left( \bar{\xi}\del h\frac{\bar{\del}^2}{{\del_-}^2}h \ +\ h\frac{\bar{\del}^2}{{\del_-}^2}(\bar{\xi}\del h)\ -\ 2\ \frac{\bar{\del}}{\del_-} (\bar{\xi}\del h) \frac{\bar{\del}}{\del_-}h\right)\ +W (\xi)\ ,\\\\
&\ =\ \mathcal{X}\ +\ \mathcal{Y}\ +\ W(\xi)\ ,
\end{split}
\ee
\ndt
and
\be 
\label{xi}
\begin{split}
W&= 2\kappa\ \xi \bar{\del} \bar{h}\ \del_-^2 \left( h\frac{\bar{\del}^2}{\del_-^2}h\ -\ \frac{\bar{\del}}{\del_-}h \frac{\bar{\del}}{\del_-} h\right)\\
& \quad +\ 2\kappa\ \bar{h}\ \del_-^2 \left( \xi \bar{\del}h \frac{\bar{\del}^2}{\del_-^2}h\ +\ h\frac{\bar{\del}^2}{\del_-^2}(\xi \bar{\del}h)\ - 2\frac{\bar{\del}}{\del_-}(\xi\bar{\del}h)\frac{\bar{\del}}{\del_-}h \right)\ , \\
&=0\ ,
\end{split}
\ee
by partial integrations (similarly, from the variation of the other cubic term we will have no $\bar{\xi}$ terms). $\mathcal{X}$ and $\mathcal{Y}$ can be further simplified using partial integrations. The results are as follows
\be 
\label{X}
\mathcal{X}\ =\ - 2\kappa\ \bar{\xi}\ \bar{h}\ {\del_-}^2 \del \left( h \frac{\bar{\del}^2}{{\del_-}^2} h\ \right) + 2\kappa\ \bar{\xi}\ \bar{h}\ {\del_-}^2 \del\left( \frac{\bar{\del}}{\del_-}h \frac{\bar{\del}}{\del_-}h \right)\ ,
\ee
and
\be \label{Y}
\begin{split}
\mathcal{Y}\ &=\ 2\kappa \bar{h}\ \bar{\xi} {\del_-}^2\ \del  \left( h\frac{\bar{\del}^2}{{\del_-}^2}h \right) \ -\ 2\kappa \bar{h}\bar{\xi} {\del_-}^2\ \del \left(\frac{\bar{\del}}{\del_-} h \frac{\bar{\del}}{\del_-}h\right)\\\\
& \qquad -\ 4\kappa\ \bar{\del} \bar{\xi}\ \frac{\del}{\del_-}h\ \frac{\bar{\del}}{\del_-}h\ 
 \del_-^2 \bar{h}\ +\ 2\kappa\ \bar{\del}^2\bar{\xi}\ \frac{\del}{\del_-^2}h\ h\ \del_-^2 \bar{h}\ +\ 4\kappa\ \bar{\del} \bar{\xi}\ \frac{\del \bar{\del}}{\del_-^2}h\ h\ \del_-^2 \bar{h}\ .
\end{split}
\ee
The first two terms in (\ref{Y}) cancel against (\ref{X}) leaving us with
\be 
\delta^{\kappa^0}(\text{cubic terms})\  =\ -4\kappa\ \bar{\del} \bar{\xi}\ \frac{\del}{\del_-}h\ \frac{\bar{\del}}{\del_-}h\ 
 \del_-^2 \bar{h}\ +\ 2\kappa\ \bar{\del}^2\bar{\xi}\ \frac{\del}{\del_-^2}h \ h\ \del_-^2 \bar{h}\ +\ 4\kappa\ \bar{\del} \bar{\xi}\ \frac{\del \bar{\del}}{\del_-^2}h\ h\ \del_-^2 \bar{h}
\ee
We now move to the quartic vertex and focus on the relevant $\kappa^{-1}$ variation. We focus on the $\bar{\xi}$ terms since the $\xi-$dependent terms may be obtained by complex conjugation. 
\be
\begin{split}
\delta^{\kappa^{-1}} (\text{quartic terms})\  = \mathcal{A}\ +\ \mathcal{B}\ +\ \mathcal{C}\ +\ \mathcal{D}\ ,
\end{split}
\ee
where
\be 
\label{A}
\begin{split}
\mathcal{A}\ &=\  -\ 4\kappa\ \bar{\del} \bar{\xi} \ \del h\ \frac{1}{{\del_-}^2}\ \left(\frac{\bar{\del}}{\del_-}h\  {\del_-}^3\ \bar{h}\ -\ h\  {\del_-}^2 \bar{\del} \bar{h} \right)\\
&=\ 4\kappa\ \bar{\del} \bar{\xi}\ \frac{\del}{\del_-}h\ \frac{\bar{\del}}{\del_-}h\ 
 \del_-^2 \bar{h}\ -\ 4\kappa\ \bar{\del}^2\bar{\xi}\ \frac{\del}{\del_-^2}h\ h\ \del_-^2 \bar{h}\ -\ 4\kappa\ \bar{\del} \bar{\xi}\ \frac{\del \bar{\del}}{\del_-^2}h\ h\ \del_-^2 \bar{h}\ ,
\end{split}
\ee
\be 
\label{D}
\begin{split}
\mathcal{D}\ &=\ -\ 2\kappa^2\ \del h\ \bar{\del}^2\bar{\xi}\ \left(\ \frac{1}{{\del_-}^2}\ (\del_-h\ \del_- \bar{h})\ -\ h\ \bar{h}\right)\\
&=\ -\ 2\kappa\ \bar{\del}^2 \bar{\xi}\ \frac{\del}{\del_-}h\ \del_- h\ \bar{h}\ +\ 2\kappa\ \bar{\del}^2 \bar{\xi}\ \frac{\del}{\del_-^2}h\ h\ \del_-^2 \bar{h}\ .
\end{split}
\ee
Notice that the terms in (\ref{A}) along with the second term in (\ref{D}) cancel the entire contribution from the cubic vertex. We now move to
\be
\label{B}
\mathcal B= +\ 2 \kappa\ \bar\del^2\bar\xi\,h\,\frac{1}{\del_-}\ (\del \bar{h}\ \del_- h\ -\ \del_- \bar{h}\ \del h)\ ,
\ee
and find that
\be 
\label{Bprime}
\begin{split}
\mathcal{B}  -\ 2\kappa\ \bar{\del}^2 \bar{\xi}\ \frac{\del}{\del_-}h\ \del_- h\ \bar{h}=+\kappa\, \bar{\del}^2\bar{\xi}\, h\,h\,\del \bar h\ .
\end{split}
\ee
Finally, we turn to the third term
\be 
\label{C}
\begin{split}
\mathcal{C}\ &=\ +\,2\kappa h\ \bar{\del}\bar{\xi}\ \left(\ \bar{\del} h\ \del \bar{h}\ +\ \del h\ \bar{\del} \bar{h}\ -\ \del_- \bar{h}\ \frac{\del \bar{\del}}{\del_-}h\ -\ \del_- h\ \frac{\del \bar{\del}}{\del_-}\ \bar{h} \right)\ \\
&=\ +2\,\kappa \, \bar{\del}\bar{\xi}\,h\, \del h \,\bar{\del} \bar h\,-\kappa\,\bar\del^2\bar\xi\,h\, h\, \del \bar h -\  2\kappa h\ \bar{\del}\bar{\xi}\ \del_- \bar{h}\ \frac{\del \bar{\del}}{\del_-}h\ .
\end{split}
\ee

\newpage

\renewcommand{\theequation}{B-\arabic{equation}}
\setcounter{equation}{0}  
\section*{Appendix {\bf {B}}}  

\vskip 0.5cm
\subsection*{Detailed computation of $\mathcal D\bar h$ at order $\kappa^2$}

\vskip 0.2cm

\ndt In this appendix, we present details of how (\ref {covd}) is derived in subsection (\ref {qfs}), starting from (\ref {hamiltonian}).

\subsubsection*{from line 3}

\vskip 0.1cm
\ndt Contribution to $\mathcal D \bar h$

\bea
&&-2\,\kappa^2\,\parm^2\bar h\,\fr{\parm^4}(\parm^2 h\partial \bar h-2\parm \partial h \parm \bar h\,)
\eea

\ndt Remaining terms 
\bea
-4\,\kappa^2\fr{\parm^2}(\parm h\parm \bar \partial \bar h)\,\fr{\parm^2}(\parm\partial h\parm \bar h)
\eea

\subsubsection*{from line 4}

\vskip 0.1cm
\ndt Contribution to $\mathcal D \bar h$

\bea
&&+2\,\kappa^2\,\parm \bar h \fr{\parm^2}(\parm h \partial \bar h -2 \partial h \parm \bar h)
\eea

\ndt Remaining terms 
\bea
-4\,\kappa^2\,\fr{\parm}(\parm h\,\bar\partial \bar h)\,\fr{\parm}(\partial h\parm \bar h)
\eea

\subsubsection*{from line 5 - I}

\vskip 0.1cm
\ndt Contribution to $\mathcal D \bar h$

\bea
&&+6\,\kappa^2\,\fr{\parm^2}(\parm h \parm \bar h)\,\partial \bar h
\eea

\subsubsection*{from line 5 - II}

\vskip 0.1cm
\ndt Contribution to $\mathcal D \bar h$

\bea
&&-6\,\kappa^2\,\parm \bar h\,\fr{\parm^2}(\parm h \partial \bar h)
\eea

\subsubsection*{from line 6}

\ndt Contribution to $\mathcal D \bar h$

\bea
&&-2\,\kappa^2\,\fr{\parm^2}(\parm h\parm \bar h)\partial \bar h \nn \\
&&+4\kappa^2\,h\bar h\,\partial \bar h \nn \\
&&+4\kappa^2\,\frac{\partial}{\parm}\{\parm \bar h\,(\fr{\parm^2}(\parm h\parm \bar h)-h \bar h)\,\}
\eea
\ndt Remaining terms 
\bea
-4\kappa^2\fr{\parm^2}(\parm h\parm \bar h)\partial h \bar\partial \bar h
\eea

\vskip 0.1cm

\ndt The ``Remaining terms", when taken together combine to yield the structures we want: that is $X\bar\del h$ or $Y\del \bar h$ which simply adds factors of $X$ or $Y$ to $\D \bar h$ or $\bD h$.


\newpage

\begin{thebibliography}{Ref}
\bibitem{ABM}{S. Ananth, L. Brink and M. Mali, {\it JHEP} {\bf 1508}, 153 (2015), arXiv:1507.01068.}
\bibitem{MH}{M. Henneaux, D. Persson, and P. Spindel, {\it Living Rev. Rel.} {\bf 11} 1 (2008),  arXiv:0710.1818.}
\bibitem{ZB2}{Z. Bern, J. Carrasco, D. Forde, H. Ita and H. Johansson, {\it  Phys. Rev.} {\bf D 77} 025010 (2008),  arXiv:0707.1035.}
\bibitem{BMS}{A. Strominger, {\it JHEP} {\bf 1407} 152 (2014),  arXiv:1312.2229.}
\bibitem{KLT}{H. Kawai, D. Lewellen and S. Tye, {\it Nucl. Phys.} {\bf B 269} (1986) 001. \\
S. Ananth and S. Theisen, {\it Phys. Lett.} {\bf  B 652} (2007) 128, arXiv:0706.1778. \\
S. Ananth, {\it Int. J. Mod. Phys.} {\bf D 19} (2010) 2379,  arXiv:1011.3287.}
\bibitem{ZB1}{Z. Bern, L. Dixon and R. Roiban, {\it Phys. Lett.} {\bf B 644} 265 (2007), hep-th/0611086}.
\bibitem{ABMj}{S. Ananth, L. Brink and S. Majumdar, {\it {JHEP}} {\bf 1603}, 051 (2016), arXiv:1601.02836.}
\bibitem{SS}{J. Scherk, J. Schwarz, {\it Gen.Rel.Grav.} {\bf 6} 537-550 (1975).}
\bibitem{BCL}{I. Bengtsson, M. Cederwall and O. Lindgren, {\it {G\"oteborg-83-55}} (1983).}
\bibitem{SA}{S. Ananth, {\it Phys. Lett.} {\bf B 664} 219 (2008), arXiv:0803.1494.}
\bibitem{ABHS}{S. Ananth, L. Brink, R. Heise, H. G. Svendsen, {\it Nucl.Phys.} {\bf B753} 195-210 (2006), hep-th/0607019.}
\bibitem{BBK}{A. Bengtsson, L. Brink and S. Kim, {\it {JHEP}} {\bf 1303}, 118 (2013), arXiv:1212.2776.}
\bibitem{MHV}{E. Witten, {\it {Commun. Math. Phys.}} {\bf {252}}, 189 (2004), hep-th/0312171. \\
A. Gorsky and A. Rosly, {\it {JHEP}} {\bf {0601}} (2006) 101, hep-th/0510111. \\
P. Mansfield, {\it {JHEP}} {\bf {0603}} (2006) 037, hep-th/0511264. \\
S. Ananth, S. Kovacs and S. Parikh, {\it {JHEP}} {\bf 1105}, 051 (2011), arXiv:1101.3540. \\
L. Dixon, arXiv:1310.5353. \\
S. Ananth, arXiv:1603.02795.} 
\bibitem{BLN}{L. Brink, O. Lindgren and B. E. W. Nilsson, {\it Nucl. Phys.} {\bf B 212}, 401 (1983).}
\bibitem{BBB}{A. Bengtsson, I. Bengtsson and L. Brink, {\it Nucl. Phys.} {\bf B 227}, 41 (1983).}
\bibitem{ABKR}{S. Ananth, L. Brink, S. Kim and P. Ramond, {\it Nucl. Phys.} {\bf B 722} 166 (2005), hep-th/0505234.}
\bibitem{BKR}{L. Brink, S. Kim and P. Ramond, {\it {JHEP}} {\bf 0806}, 034 (2008), arXiv:0801.2993.}
\bibitem{BT}{L. Brink and A. Tollsten, {\it Nucl. Phys.} {\bf B 249} (1985) 244.}
\end{thebibliography}
\end{document}